# TIME-DEPENDENT THEORY OF SOLAR MERIDIONAL FLOWS

James H. Shirley

Jet Propulsion Laboratory, Pasadena, CA USA

James.H.Shirley@jpl.nasa.gov

# ABSTRACT

We explore consequences for the solar dynamo of a newly-developed physical hypothesis describing a weak coupling of the orbital and rotational motions of extended bodies. The coupling is given by - $c$ ($\dot{\boldsymbol{L}} \times \boldsymbol{\omega_a}$) $\times \boldsymbol{r}$, where $\dot{\boldsymbol{L}}$ represents the rate of change of barycentric orbital angular momentum, $\boldsymbol{\omega_a}$ is the angular velocity of rotation, $\boldsymbol{r}$ is a position vector identifying a particular location in a coordinate system rotating with the Sun, and $c$ is a coupling efficiency coefficient. This form of coupling has no dependence on tides. The coupling expression defines a non-axisymmetric global-scale acceleration field that varies both in space and with time. Meridional components of acceleration typically dominate in equatorial and middle latitudes, while zonal accelerations become increasingly significant at higher latitudes. A comparison of the waveform of the putative dynamical forcing function with the time series for measured solar meridional flow speeds from Sunspot Cycle 23 yields correlations significant at the 99.9% level. We introduce the possibility of a destructive interaction between the predicted large-scale flows (due to orbit-spin coupling) and the dynamo mechanism(s) of the 22-year magnetic activity cycle; observations in recent cycles of higher meridional flow speeds during episodes of reduced solar sunspot activity may be explained as a consequence of differences between the phasing of the magnetic cycle and the phase of the forcing function. Algorithms are provided for calculating meridional and zonal orbit-spin coupling accelerations within the Sun as a function of latitude, longitude, depth, and time.

# 1. INTRODUCTION

The past two decades have witnessed a rapid growth in our understanding of the physical mechanisms underlying the 22-year Hale solar magnetic cycle (Charbonneau, 2014; Cameron et al., 2016). Improved observational methods and techniques, and advancements in numerical modeling of the solar dynamo, have each made important contributions to the growth of knowledge in this area. The interplay between modeling and observations, together with advancements in computational capabilities, has focused attention on a small number of critical topics and questions. In particular: *Babcock-Leighton flux transport dynamos* (Choudhuri et al., 1995; Durney, 1995; Dikpati & Charbonneau, 1999; Dikpati & Gilman, 2001; Wang et al., 2002; Nandy & Choudhuri, 2002; Chatterjee et al, 2004) are now considered to represent a leading explanatory framework for the solar magnetic cycle (Karak, 2010; Dikpati & Gilman, 2012; Choudhuri & Karak, 2012; Upton & Hathaway, 2014; Lemerle et al., 2015), as these can account for many of the observed features of the solar magnetic cycle. *Polar fields* play a pivotal role, in flux transport dynamos, by supplying the flux needed for the regeneration of the toroidal field components (Wang et al., 1989, 2002a, 2002b, 2009; Ulrich & Boyden, 2005; Janardhan et al., 2010; Dikpati, 2011; Cameron & Schüssler, 2015; Wang, 2016). This crucial role is well supported by observations, as polar field strengths following solar sunspot cycle maxima are strongly correlated with sunspot cycle amplitudes in following cycles (Schatten et al. 1978; Schatten & Hedin, 1984; Schatten & Sofia, 1987; Schatten, 2003, 2005; Svalgaard et al., 2005).

As with the polar fields, *large-scale meridional flows* of solar materials, which are a primary focus of the present paper, are widely considered to play an important role in flux transport dynamos. Meridional flows are thought to 'subduct' the surface poloidal flux in polar regions, thereafter conveying this equatorward along the base of the convection zone to lower latitudes. This transport 'closes the loop' and allows the regeneration of magnetic fields associated with sunspots (Choudhuri et al., 1995; Durney, 1995; Dikpati & Charbonneau, 1999; Wang et al., 2002; Nandy & Choudhuri, 2002; Chatterjee et al, 2004; Jiang et al., 2010; Dikpati & Gilman, 2012). Models indicate that meridional flow speeds may play a central role in the determination of sunspot cycle periods, shapes, and amplitudes (Wang & Sheeley, 1991; Choudhuri et al., 1995; Durney, 1995; Dikpati & Charbonneau, 1999; Charbonneau & Dikpati, 2000; Hathaway et al., 2003; Basu & Antia, 2003; Chatterjee et al, 2004; Karak, 2010; Dikpati & Gilman, 2012; Choudhuri & Karak, 2012; Dikpati & Anderson, 2012).

Observations of solar meridional flows (Duvall, 1979; Ribes et al., 1985; Snodgrass & Dailey, 1986; Wang et al., 1989; Komm et al., 1993; Hathaway et al., 1996; Giles et al., 1996; Braun & Fan, 1998; Chao & Dai, 2001; Javaraiah & Ulrich, 2006; Ulrich, 2010; Hathaway & Rightmire, 2010; Komm et al, 2015a) are broadly consistent with the patterns of large scale solar meridional flows employed in numerical modeling. Many investigations document variations of meridional flow speeds that appear to be linked with the 11-year (Schwabe) sunspot cycle (Komm et al., 1993, 2011, 2015a; Snodgrass & Dailey, 1996; Meunier, 1999; Chou & Dai, 2001; Hathaway et al., 2003; Basu & Antia, 2003, 2010; Javaraiah & Ulrich, 2006; Hathaway & Rightmire, 2010, 2011; Ulrich, 2010; Hathaway & Upton, 2014, 2016; Zhao et al., 2014; Hazra et al., 2015). Such evidence confirms the importance of the role of meridional flows as a key component of the underlying mechanism(s) responsible for the solar magnetic cycle.

While a number of early flux transport dynamo models employed meridional flows with constant flow speeds (*cf*. Choudhuri et al, 1995; Dikpati & Gilman, 2001; Nandy & Choudhuri, 2002), it has long been apparent that these large-scale flows must vary with time, in order to more adequately account for the phenomena of the solar magnetic cycle. Stochastic variability of flow speeds has been employed in some investigations (*cf.* Charbonneau & Dikpati, 2000; Usoskin et al. 2009; Nandy et al., 2011; Hazra et al., 2015), while in many other studies, flow speed changes have been specified externally as a feature of the experimental design (Dikpati & Charbonneau, 1999; Wang et al., 2002a, 2002b, Karak, 2010; Dikpati & Anderson, 2012; Dikpati & Gilman, 2012; Upton & Hathaway, 2014; Hathaway & Upton, 2016). In still other studies, the problem has been inverted, with solar activity levels being used to estimate past solar meridional flow speeds (Passos & Lopes, 2008). Despite abundant evidence of temporal variability of meridional flow speeds on and within the Sun, the origins of this variability remain unknown.

Two further questions of some importance involving solar meridional flows may be noted in passing. First is the observed negative correlation between measured solar meridional flow speeds and sunspot cycle activity levels (Komm et al., 1993, 2015a; Chao & Dai, 2001; Basu & Antia, 2003, 2010; Hathaway & Rightmire, 2010; Hathaway & Upton, 2016). This finding seems to conflict with expectations from some numerical modeling investigations, in which faster flows may allow shorter times for flux diffusion and thereby engender shorter cycle periods and higher levels of activity. Secondly, as observational techniques and processing methods have improved, evidence has accumulated for the intermittent presence of *multiple* meridional flow cells, separated both in depth and in latitude, at different times (Haber et al., 2002; Ulrich, 2010; Zhao et al., 2013; Schad et al., 2013; Kholikov et al., 2014). Clearly the single-cell-per-hemisphere 'conveyor belt' model for meridional flows is insufficiently detailed and must be updated and upgraded. Attempts to model more complex meridional flow patterns have met with some success (Jouve & Brun, 2007; Dikpati, 2014; Hazra et al., 2014; Belucz et al., 2015).

Considerations such as these led Belucz et al. (2013) to call for "time-dependent theories that can tell us theoretically how this circulation may change its amplitude and form in each hemisphere." Such a theory now exists for planetary atmospheres (Shirley, 2017). The dynamical orbit-spin coupling hypothesis of Shirley (2017) has been evaluated through numerical modeling of the atmospheric circulation of Mars (Mischna & Shirley, 2017) using the MarsWRF general circulation model (GCM; Richardson et al., 2007; Toigo et al., 2012). The GCM, with fully deterministic orbit-spin coupling, was able to reproduce atmospheric conditions favorable for the occurrence of perihelion-season global-scale dust storms on Mars, within the model years in which such storms were actually observed. The physical hypothesis is similarly applicable to the case of the circulation of the solar convective zone, although both the catalog of forces and the physical interactions are quite different.

The primary objectives of this paper are to describe the theoretical basis and observational support for orbit-spin coupling as a factor that may contribute significantly to the time-variability of large-scale flows within the Sun. As with Mars, the physical hypothesis is susceptible to evaluation through numerical modeling. A third objective is to introduce and describe the algorithms that will allow such modeling to go forward.

In Section 2 we highlight an unexplained correlation that appears to link the 22-year Hale cycle of solar magnetic activity with dynamical processes occurring within the solar system (Jose, 1965). The discussion of this section bears on the open question of what may cause the solar dynamo to emerge from Maunder Minimum conditions. Following this introduction, the physical hypothesis is outlined in Section 3, while in Section 4 some of the predictions of the hypothesis are compared with past solar observations. The new perspective afforded by the physical hypothesis allows us to review and offer possible interpretations for some of the unusual phenomena and events of Solar Cycle 23.

In Section 5 we compare the record of solar meridional flow speeds of small magnetic features of Hathaway & Rightmire (2010) with the waveform of the putative orbit-spin coupling forcing function for the same years, obtaining a correlation significant at the 99.9% level. Section 6 reviews and interprets a number of the events and phenomena of Cycle 24, and includes a short discussion of the issue of the observed anti-correlation of meridional flow speeds and sunspot cycle activity levels.

We recognize that the problem of the variability of meridional flows within the Sun is unlikely to be resolved through the addition to our dynamo models of a single new kinematic forcing function. The backreaction due to magnetic forces must certainly enter the problem; these must evidently be modeled in parallel (Shibahashi, 2004; Rempel, 2006, 2007; Passos et al., 2012; Beaudoin et al, 2013; Dikpati, 2014; Hotta et al., 2015; Karak & Cameron, 2016; Guerrero et al., 2016). The problem of system memory remains an open question of considerable importance (Yeates et al., 2008; Dikpati et al., 2010; Dikpati & Anderson, 2012; Muñoz-Jaramillo et al., 2013), and within this problem area, the relative roles played by advection and diffusion (Dikpati et al., 2006; Choudhuri et al., 2007) must likewise demand modelers' attention. We consider these topics, and others, in the discussion of Section 7.

The orbit-spin coupling hypothesis leads naturally to a new interpretation for the observed temporal variability of solar meridional flows. Meridional flow cells may evidently either strengthen, or weaken and decay, with time. Strengthening meridional flow cells are here tentatively identified as temporary repositories for the momentum gained during a transfer of angular momentum between the orbital 'reservoir' and the reservoir of the solar rotational motion. The accelerations due to orbit-spin coupling may, at other times, generate patterns of large-scale flows that interfere destructively with meridional flow cells already in existence.

Conclusions are presented in Section 8, while algorithms for calculating the orbit-spin coupling accelerations are found in the Appendix to this paper.

## 2. AN UNEXPLAINED CORRELATION

Conditions and processes within the Sun leading to the occurrence of grand minima of solar activity are a long-standing topic of interest (Eddy, 1976; Wang & Sheeley, 2003; Choudhuri & Karak, 2009; Usoskin et al. 2009, 2015; Karak, 2010; Feynman & Ruzmaikin, 2011; Vaquero et al., 2011; McCracken & Beer, 2014; Zolotova & Ponyavin, 2014). The most recent such event, the Maunder Minimum, extended from approximately 1645 CE to 1715 CE (Eddy, 1976; Stuiver & Quay, 1980). The uppermost pair of curves of Fig. 1 represents Hale-cycle sunspot numbers for the 60-year period beginning in 1710, together with the rate of change of the orbital angular momentum of the Sun (with respect to the solar system barycenter) for the same years. The correlation coefficient $r$ obtained through a comparison of these waveforms is 0.72. The relationship displayed was first noted by Jose (1965), who also recognized and described a repetition cycle of ~178 years characterizing the waveform of $d\boldsymbol{L}/dt$.

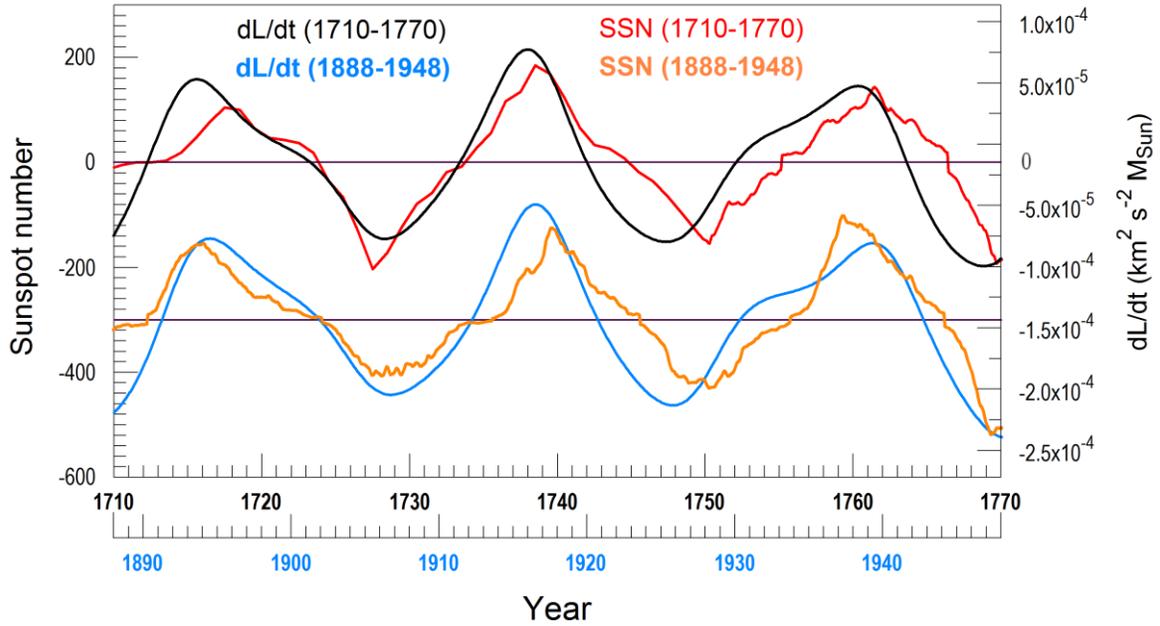

**Figure 1**. Hale cycle sunspot numbers and the rate of change of the Sun's orbital angular momentum *dL/dt*, for the intervals from 1710 to 1770 (upper panel) and from 1888 to 1948 (lower panel; both after Jose, 1965). International yearly (1710-1749) and 13-month smoothed monthly mean sunspot numbers (1750-present) are from the Sunspot Index and Long-term Solar Observations (SILSO) compilation (Clette et al., 2014). The unit of mass for the dynamical waveform (with scale at right) is the solar mass (*M_Sun*). The paired curves for the later time interval have been offset by -250 SSN for clarity. As in Jose (1965), the dynamical waveform (*dL/dt*) includes contributions to the solar motion from the giant planets only, which together account for > 98% of the angular momentum of the solar system. Methods for obtaining the dynamical waveform resulting from giant planet contributions only are described in Jose (1965) and in Fairbridge & Shirley (1987).

Also shown in Figure 1 are the corresponding waveforms of the 22-year Hale magnetic cycle in sunspot numbers and *dL/dt* for the years 1888-1948 (178 years later). We first note the striking similarity of the waveforms for $dL/dt$, for these two periods, which is a consequence of solar system dynamics. We next note, for the second set of curves, a similar correlation of sunspot numbers with $dL/dt$ ($r$=0.85 for this interval). We need not be concerned with establishing statistical significance for the above relationships, as this has already been accomplished, for the entire period since 1700, by Paluš et al. (2007). Visual comparison of the two curves representing sunspot numbers reveals a number of broad similarities. Similar drifts in the phasing of the compared waveforms are seen, for instance near 1750 (1928) and 1770 (1948). While a correlation such as that shown for 1710-1770 might plausibly arise as a result of random variability within physically unrelated time series, it severely strains credibility to assert that precisely the same sort of relationship should randomly arise for the second interval as well.

In the present paper we are principally concerned with identifying and characterizing physical processes that may potentially account for such relationships. We must immediately reject the possibility that gravitational tides might give rise to these correlations; the waveform of $dL/dt$ (as illustrated in Fig. 1) is principally determined by the positions and motions of the giant planets, and of these, only Jupiter gives rise to an appreciable tide. Furthermore, the tidal acceleration of Jupiter is smaller than the acceleration of gravity at the solar surface by a factor of ~$10^{12}$; and the tide thus raised has an amplitude of only ~ 1 mm. In Section 3.2 we will compare and contrast quantitative aspects of the planetary tidal and orbit-spin coupling hypotheses in somewhat more detail.

Solar phenomena of the two periods illustrated in Fig. 1 share a number of other similarities. In each case, the amplitude of the first 11-year (Schwabe) cycle following the intervals shown was quite large; the two highest-amplitude Schwabe cycles on record are those of 1957 and 1778. While we do not have modern observational data for the earlier period, we know that the solar coronal magnetic field strength approximately doubled in the first half of the 20th century (Lockwood et al., 1999). Given the large amplitude of the sunspot cycle maximum of 1778, it would not be surprising to find that the same was true for the earlier period. By inspection of Fig. 1, it is also noteworthy that double-peaked Schwabe cycles are poorly represented in these intervals. Finally, Benestad (2005) has pointed out a striking stabilization of 'solar cycle length characteristics' beginning at the turn of the 20th century, in comparison to prior decades. Our earlier period (1710-1770) presumably shares this characteristic (though this question will not be pursued further here).

If the correspondence shown in Fig. 1 persisted throughout the historic record, there would be little room for controversy concerning the hypothesis of a physical relationship between these indices. However, periods outside these intervals are characterized by more dramatic and irregular phase and amplitude modulation of the $dL/dt$ waveform, and the illustrated correspondence in phase is less in evidence. As will be shown in Section 4, this does not constitute a disqualifying argument for the hypothesis of a physical relationship; instead, this provides us with valuable information on the nature of possible physical mechanisms, with specific implications for the operation of the solar dynamo.

The intervals chosen for display in Fig. 1 were not selected arbitrarily, but instead represent previously recognized periods in which the Sun's barycentric orbital motion was relatively regular and well ordered (Charvátová & Střeštík, 1991; Charvátová & Hejda, 2014). During the past millennium, the Sun has spent about one-third of the time under dynamical conditions closely similar to those of Fig. 1. We opened this Section by mentioning that the interval from 1710-1770 is notable for being a period in which the Sun was emerging from Grand Minimum conditions. Taken at face value, the relationships of Fig. 1 suggest that, during this interval, some physical factor linked with $d\boldsymbol{L}/dt$ may have in some way "entrained" the solar dynamo, in the process resonantly pumping up the level of magnetic activity. If so, then this could represent a suitable mechanism for emerging from prolonged sunspot minimum conditions. Dynamical conditions similar to those illustrated in Fig. 1 also occurred contemporaneously with the Sun's emergence from the Wolf Minimum, which ended near 1350 CE, and the Spörer Minimum, which ended near 1540 CE (Charvátová & Hejda, 2014, Fig. 1).

While the possibility of an entrainment of the solar dynamo by some physical factor linked with $d\boldsymbol{L}/dt$ was recognized more than half a century ago (Jose, 1965), further progress has been greatly impeded by the lack of a viable physical mechanism. As documented more comprehensively in Shirley (2017), the only known forces arising in the solar system dynamical environment capable of imparting differential motions to the constituent particles of extended bodies are the tide-raising forces; and, as already noted, the tides experienced by the Sun are both quantitatively insufficient and qualitatively inadequate to account for the relationships displayed in Fig. 1.

## 3. THE ORBIT-SPIN COUPLING HYPOTHESIS

No attempt to survey the scientific literature treating planetary theories of solar activity will be made here. However, for completeness, we will briefly trace the evolution of the concept of a coupling of orbital and rotational motions as it relates to the phenomena of Fig. 1. The hypothesis appears to have emerged in conceptual form in the late 1970s (Blizard, 1981, and references therein). It was subsequently invoked and discussed in a number of studies by Landscheidt (1988, 1999), and by Zaqarashvili (1997), Juckett (2000, 2003), Javaraiah (2003, 2005), Wilson (2008, 2013), and Wolff & Patrone (2010). Three of these studies presented quantitative physical hypotheses that could readily be evaluated; these were the studies by Zaqarashvili (1997), Juckett (2003), and Wolff & Patrone (2010). The hypothesis of Wolff & Patrone (2010) was later explored and evaluated in a modeling study by Cionco & Soon (2015), who noted some agreement of the tested parameter to the timing of past solar prolonged minima, but found little evidence of significant effects on the time scales of the Schwabe cycle. The innovative, potentially groundbreaking coupling mechanisms described in Zaqarashvili (1997) and in Juckett (2003) were later called into question by Shirley (2006), who pointed out an inappropriate use of rotational equations by those authors for physical problems involving orbital revolution.

Recent developments in this area have arisen not in connection with additional solar investigations but from attempts to understand the physical origins of the interannual variability of the atmospheric circulation of Mars. Shirley (2015) uncovered and described a small number of remarkably systematic relationships linking the occurrence and non-occurrence of global-scale dust storms on Mars with the variability of that planet's orbital angular momentum with respect to the solar system barycenter. An entirely new approach to the orbit-spin coupling problem was subsequently formulated (Shirley, 2017; hereinafter, Paper 1, or P1). Generalized predictions based on the physical hypothesis of P1 have been tested both by statistical methods (Shirley & Mischna, 2017) and through numerical modeling investigations employing a Mars atmospheric general circulation model (Mischna & Shirley, 2017). The present investigation benefits greatly from lessons learned during these prior efforts.

The somewhat lengthy formal derivation of Paper 1 will not be repeated or further summarized here. Instead we will immediately turn our attention to the result of that derivation, which for convenience has been labeled the "coupling term acceleration," or *CTA*:

$$CTA = -c \left( \dot{L} \times \omega_\alpha \right) \times r \quad (1)$$

Here $\dot{L}$ represents the rate of change of barycentric orbital angular momentum, $\omega_\alpha$ is the angular velocity of rotation, $r$ is a position vector identifying a particular location in a rotating coordinate system, and $c$ is a scalar coupling coefficient. $\dot{L}$, or $dL/dt$, is the dynamical variable plotted in juxtaposition with sunspot numbers in Fig. 1. The vector $\omega_\alpha$ lies along the axis of rotation of the subject body, and the coupling efficiency coefficient $c$ is a variable that is to be determined through numerical modeling. The origins, nature, and likely value of $c$ (for the case of Mars) are discussed extensively in Paper 1. The likely value of $c$ for most solar system cases is constrained by observations to be quite small. We will return to a consideration of the nature and numerical value of $c$ for the solar case following a short description of the spatial geometry and temporal variability of the *CTA* as specified by Equation (1).

### 3.1. Spatiotemporal variability of the CTA

The global pattern of accelerations identified and described by Equation (1) is illustrated in Fig 2. The origins of this pattern may be understood in the following way. The cross product of $\dot{L}$ and $\omega_\alpha$ is a vector that lies within the equatorial plane of the subject body. Crossing this with a position vector (i.e., the radius vector $r$) yields an acceleration vector that is tangential to a spherical surface of radius $r$. The accelerations disappear in locations where $r$ is parallel to the cross product of $\dot{L}$ and $\omega_a$; one such location is seen at the lower right of the figure. Maximum values of the acceleration are found on the great circle of longitude that is 90° removed from these zero points. Exclusively meridional accelerations encircle the subject body on this great circle. This 'overturning' characteristic bears some resemblance to a classical mechanical couple, such as a belt and pulley system (P1).

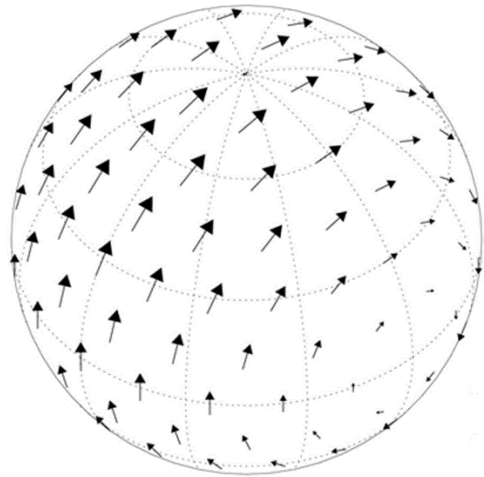

Figure 2.   Vector representation of the coupling term acceleration (*CTA*; Equation (1)) over the surface of an extended body, from a viewpoint situated above 45° N latitude. The lengths of the displayed vectors are proportional to their magnitude. Latitude and longitude grid lines at 30° intervals are shown for reference. (The acceleration vector for the north polar location has been omitted).

As the cross product of $\dot{L}$ and $\omega_a$ generally varies slowly with time (recall the time scale of Fig. 1), we must visualize the axial rotation of the Sun carrying particular locations on the Sun's surface through this spatial pattern of accelerations over the period of one sidereal rotation. (That is, for short intervals of time, the field may be considered to be approximately 'fixed in space').  For a location at 60° N, for instance, the direction of the acceleration will cycle completely in azimuth over one solar rotation, while varying systematically in magnitude over this cycle.  With respect to the rotational velocity vector for a particle at 60° N, this implies that over about half of that small circle of latitude, the acceleration will constructively add to that velocity; while in the opposite hemisphere, on the same small circle of latitude, the particle velocity associated with rotation will be correspondingly diminished by the acceleration.

Two other features of the acceleration field of Fig. 2 are worthy of note.  As just indicated, in the northern mid-latitudes of the displayed subject body, the zonal accelerations would act to augment the velocity due to axial rotation; but at the corresponding latitudes of the southern hemisphere, on the same side of the body, the opposite tendency will prevail.  While detailed modeling will be required to draw firm conclusions, this pattern seems well suited to the generation of hemispheric asymmetries of the circulations of solar materials.  As emphasized by Lemerle et al. (2015), the Coriolis force is presently the main physical factor that breaks the axial symmetry of proposed driving mechanisms for solar dynamo models, thereby allowing these to escape the fatal consequences of Cowling's theorem (Cowling, 1934).  The *CTA* may provide another source of non-axisymmetric motions of relevance to the dynamo problem.

One final point of interest relating to Fig. 2 concerns the apparent dominance of meridional components of acceleration in equatorial and middle latitudes, in comparison with the zonal components.  This disparity was evaluated quantitatively in Paper 1, where it was found that the integral of the absolute values of the magnitudes of the meridional components over the entire surface is roughly twice as large as the corresponding sum for zonal components.

To this point we have mainly considered the geometric or spatial variability of the acceleration resulting from the rotation of the subject body through the pattern (or *field*) illustrated in Fig. 2. We now turn to a consideration of the temporal variability of the field itself. From Equation (1) we recognize that $dL/dt$ (i.e., $\dot{L}$) is the principal source of variability for the *CTA*; for this reason, we will often characterize $dL/dt$ as the 'forcing function' for the *CTA*. The variability with time of $dL/dt$ has already been introduced and described in connection with Fig. 1. (To maintain consistency with previous studies, $dL/dt$ will be used in preference to the overdot notation as a label for the forcing function within this paper).

The reversal in sign of the $dL/dt$ waveform at intervals of 10-12 years (Fig. 1) is one of the most significant features of the dynamical forcing function with respect to the solar dynamo problem. A change of sign of the $dL/dt$ waveform dictates a complete reversal in the directions of the acceleration vectors of Fig. 2. A reorganization of pre-existing large-scale flow patterns may possibly result from this reversal in the directions of the accelerations. It is also noteworthy that the transitional periods, adjacent to and including the zero-crossing times of the $dL/dt$ waveform, are intervals when the putative forcing by the coupling term accelerations must necessarily diminish and disappear. We must thus envision a pulsation, characterized by growth, decay, and reversal, of the magnitudes of the vectors making up the acceleration field of Fig. 2, as a result of the variability with time of the $dL/dt$ waveform.

For ease of reference, both here and in P1, we refer to time intervals proximate to the zero-crossing times of the $dL/dt$ waveform (Fig. 1) as 'transitional intervals.' Similarly, both here and in P1, we label the times when the $dL/dt$ waveform is greater than zero as 'positive polarity intervals,' while the label 'negative polarity intervals' refers to times when the rate of change is negative. In prior investigations with the Mars atmosphere (Mischna and Shirley, 2017), the positive and negative polarity episodes were typically found to be accompanied by significant differences in the morphology of the meridional overturning (or "Hadley") circulation for Mars (as a function of season). We thus expect it will similarly be of interest to assess the solar response to the polarity changes and changes of magnitude of the forcing function; these are questions best addressed by numerical modeling, preferably in three spatial dimensions plus time.

### 3.2. Magnitude of the CTA for the case of the Sun

A discussion of the constraints from astronomical observations and from solar system ephemeris calculations that bear on the permissible levels of angular momentum exchange between orbital and rotational reservoirs within the solar system is found in Section 5 of Paper 1. There it was found that an ongoing exchange between orbital and rotational reservoirs of angular momentum of more than a few parts in $10^{-12}$ of the orbital component (for the most accurately observed terrestrial planets) could most probably be ruled out by observations. From this and other considerations, it was determined that the efficiency of any such transfer must be extremely low. The leading coefficient $c$ of Equation (1) was introduced to enable quantitative assessment of the efficiency of the putative coupling. A zero value of $c$ would imply that no exchange of angular momentum is occurring; while a non-zero value, if found, might convey important information regarding both the nature of the coupling mechanism and the response of the subject body. As noted in P1, determination of $c$ from first principles is an intractable problem, at least at present, as $c$ represents a placeholder for a multitude of possible interactions, some of which may be dissipative. The multiple concentric shells of extended bodies, comprised of materials of differing physical properties, may exhibit responses to forcing that differ widely. These responses may additionally vary as a function of frequency or temperature. Thus $c$, like the specific dissipation function, $Q$, of geophysics, presently represents a catch-all parameter that in many cases may best be determined empirically.

The orbital angular momentum $\boldsymbol{L}$ of the Sun ranges between 0 and about 4.8 x $10^{40}$ kg m$^2$ s$^{-1}$, except in rare instances of retrograde solar motion (*cf.* Javaraiah, 2005); in these cases the numerical value of $\boldsymbol{L}$ may be *negative* for intervals of up to about 2 years. The *rotational* angular momentum of the Sun, by comparison, is larger by a factor of 25 or more, at ~1.1 x $10^{42}$ kg m$^2$ s$^{-1}$. This quantity, while large, is considerably smaller than the total angular momentum of the solar system, which is >3.15 x $10^{43}$ kg m$^2$ s$^{-1}$. Most of the angular momentum of the solar system resides in the giant planets, with Jupiter and Saturn accounting for 86% of the total.

The following sample calculation, using input data corresponding to a date in 1986, can illustrate a typical situation. Using the methods detailed in the Appendix, we first acquire ephemeris data from the JPL Horizons System (Giorgini et al., 1996; Georgini, 2015) to obtain $d\boldsymbol{L}/dt$ as a function of time, referenced to the J2000 ecliptic coordinate system. We obtain Cartesian components (*x, y, z*) of $d\boldsymbol{L}/dt$ equivalent to [-5.569x$10^{-6}$, -3.965 x $10^{-6}$, -2.062 x $10^{-4}$]. Because the underlying ephemerides resolve distances in km, and because two factors of the orbital radius $R$ are present in the calculation of $\boldsymbol{L}$, the distance unit for the above vector components is km$^2$; further, as in Fig. 1, the unit of mass is the solar mass of 1.99 x $10^{30}$ kg.

We next resolve the vector angular velocity of the mean solar sidereal rotation $\boldsymbol{\omega_\alpha}$ (employing rotational elements of the Sun from Beck & Giles, 2005) in the same Cartesian coordinate frame, obtaining components of [3.422 x $10^{-7}$, -1.013 x $10^{-7}$, 2.843 x $10^{-6}$] (all in radians per second). The cross product of the above two vectors yields $\dot{\boldsymbol{L}} \times \boldsymbol{\omega_\alpha}$ = [-3.215 x $10^{-11}$, -5.469 x $10^{-11}$, 1.922 x $10^{-12}$]. For an approximate calculation of (maximum) magnitude, we can employ the resultant of these (i.e., 6.347 x $10^{-11}$) to obtain the product of this with $\boldsymbol{r}$ (for a location on the Sun's surface), using a value of $r = 6.96 \times 10^8$ m. We obtain a value of 4.418 x $10^{-2}$ as a result of these operations. Finally, to obtain SI units, we must multiply this number by a factor of $10^6$ to account for our use of km$^2$ units for distance. We thus obtain a value of 4.418 x $10^4$ (m$^2$ s$^{-2}$) for the overturning torque of the couple given by Equation (1) when $c = 1$.

As discussed in Section 3.5 of Paper 1, this quantity has the temporal dimension of inverse seconds squared, corresponding to an acceleration parallel to the Sun's surface, at some locality of interest. It is more than 2 orders of magnitude larger than the acceleration of gravity at the Sun's surface (~270 m s$^{-2}$), and is larger than the tidal acceleration of Jupiter by 14 orders of magnitude.

The above calculation assumes a coupling efficiency of unity, and as we already know (P1) this cannot be the case. As a practical matter, for numerical modeling purposes, we will wish to determine whether there exists a non-zero value for the coupling coefficient $c$ that will enable us to obtain an improved correspondence between numerical modeling outcomes and observations. A small number of iterations of the modified MarsWRF GCM (Mischna & Shirley, 2017), with test values of $c$ varying over several orders of magnitude, was sufficient to usefully constrain the value of $c$ for that case. An initial strategy for the Sun might be to begin with $c = 1.0 \times 10^{-13}$, and to thereafter progressively increase the $c$ value in steps of an order of magnitude from that threshold. This would initially yield accelerations about an order of magnitude larger than those of the tidal acceleration of Jupiter, with each successive iteration thereafter producing accelerations larger by a factor of 10. An upper bound for $c$ may presumably be found through a comparison of model outcomes with observations. Unrealistically large or otherwise pathological model-generated values for meridional flow speeds may for instance be obtained when the specified value of $c$ is too large.

Further discussion of the physical origins and meaning of *c* is found in P1 and also below in Section 5.  However, before moving forward, we must clarify one key factor pertaining to the above calculation, as this could potentially lead to misunderstandings.  The value of *c* as represented in Equation (1) identifies the decimal fraction of the *difference* of the orbital angular momentum as a function of time (conceptually, $\Delta \boldsymbol{L}$) that may be participating in the angular momentum exchange process.  The value thus obtained does not directly correspond to a fractional value for the *total* angular momentum ($\boldsymbol{L}$) of the subject orbital motion.

### 3.3. Predictions

A small number of predictive statements pertaining to the orbit-spin coupling hypothesis outlined above were identified and listed in Paper 1.  In that paper, the accelerations were assumed to represent a perturbing influence on large-scale circulations of planetary atmospheres that are established and maintained by other means (i.e., principally by the seasonal cycles of solar insolation, for the terrestrial planets).  From that perspective, the most fundamental consequence then foreseen centered on the likely occurrence of constructive and destructive interference effects (on the pre-existing circulation), resulting from the introduction of the *CTA* within the subject atmosphere.  This generalized outcome is somewhat difficult to isolate and quantify.  However, its recognition led to the formulation of two further predictive statements of somewhat greater utility.

With respect to numerical modeling outcomes for tests comparing circulation model runs performed both with and without the *CTA*, assuming non-negligible values for *c*:  A broader range of variability of system behaviors (for the accelerated atmosphere) was identified in P1 to represent a verifiable and testable outcome.

A third expected consequence from P1 of the putative existence of orbit-spin coupling accelerations in a planetary atmosphere is directly applicable to the solar case. Cycles of intensification and relaxation of *CTA*-accelerated planetary atmospheres were recognized as a likely outcome of orbit-spin coupling in P1, with phasing linked directly with the cyclic variability of the forcing function *dL/dt*. Under the hypothesis of Paper 1, the extrema of the *dL/dt* waveform are identified with intervals of circulatory intensification, as the predicted accelerations attain maximum values at those times. Conversely, during transitional intervals, as when the *dL/dt* waveform transits the zero line of Fig. 1, the hypothesis predicts that the affected circulation must "relax" (to an unforced state). In Fig. 1, the twelve displayed (Schwabe) sunspot cycle maxima would thus each be associated with periods of circulatory intensification. Conversely, the intervening (Schwabe) sunspot cycle minima each approximately coincide in time with intervals when the accelerations of the CTA diminish in magnitude, disappear, and re-emerge with reversed sign.

An important caveat (among several identified in P1) concerns possible effects and consequences of *system memory*. The thin Mars atmosphere, with a thermal time constant of only ~2 days, and no large thermal reservoirs comparable to Earth's oceans, has an almost-negligible system memory. Mars was thus recognized (Mischna & Shirley, 2017) as an optimal candidate for initial testing of the orbit-spin coupling hypothesis. System memory effects may be much more significant in the case of the Sun. We will return to this question in Sections 6 and 7 below.

A second important caveat from P1 concerns the interpretation of the terms 'circulatory intensification' and 'relaxation.' The application to an extended body atmosphere of the acceleration field of Fig. 2 is unlikely to produce a strictly linear response on the part of that atmosphere, due not only to the spatiotemoral variability of the forcing, but also due to non-linear interactions within the atmospheric system itself.  Morphological changes of circulatory flows may be engendered which locally interfere destructively with pre-existing flow patterns.  Velocities of large-scale flows may typically increase during intensification intervals, but we may be required to resolve these on a global scale in order to verify this.  Numerical modeling will be required in order to better identify the outcomes of the predicted cycles of 'circulatory intensification and relaxation' on and within the Sun.

### 3.4. Additional resources

The abbreviated description of the orbit-spin coupling hypothesis provided here has a number of shortcomings.  Several related topics of importance are discussed in greater detail in Paper 1.  Paper 1, for instance, briefly reviews and summarizes current knowledge in the areas of previously known coupling mechanisms and of gravitational tides and dissipative processes.  The more theoretical topic of the assumed independence of orbital and rotational motions is discussed there in some detail.  More extended discussions of the origins and nature of the coefficient $c$, and of the permissible levels of angular momentum exchange within the solar system, are also to be found in P1.  The numerical modeling investigation of Mischna & Shirley (2017) and the statistical comparisons of Shirley & Mischna (2017) also provide insights into the topic of orbit-spin coupling that go beyond the brief discussion provided here.

4.  RECENT SOLAR VARIABILITY AND ORBIT-SPIN COUPLING (1):

The orbit-spin coupling hypothesis postulates an exchange of angular momentum between the separate reservoirs of the solar orbital angular momentum and the solar rotational angular momentum. A non-negligible exchange between these reservoirs must necessarily be accompanied by changes in the rotation state of the Sun. Thus the topic of the observational record of the rotational variability of the Sun is clearly of significance for the orbit-spin coupling hypothesis.

The existence of an intimate relationship linking the variability of the solar rotation with solar activity is by now widely accepted (Zhao & Kosovichev, 2004; Georgieva et al., 2005; Rempel, 2007; Komm et al., 2011; DeRosa et al., 2012; McIntosh et al., 2014; Cameron et al., 2016). However, the possible existence of a relationship linking either of these phenomena with solar system dynamical processes, as posited here, has not received a great deal of attention. To provide context for discussions to follow, we now briefly review some key prior findings in these areas.

### 4.1. Solar rotation and the solar magnetic activity cycle

The possibility of a relationship of the solar differential rotation and the magnetic activity cycle was unproven and controversial until about 1975. A number of studies soon thereafter found evidence for such a relationship (Howard, 1976; Antonucci & Dodero, 1976; Stenflo, 1977; Clark et al., 1979). Further investigation in this area led to the discovery and recognition of the solar *torsional oscillations* (Howard & LaBonte, 1980; Howard, 1981; Snodgrass & Howard, 1985; Bogart, 1987), which are alternating bands of faster and slower (zonal) rotational motions that move both toward the equator and toward the polar regions. These zones move in wavelike fashion and flank the zones of emergent solar activity (i.e., the active regions) in solar middle latitudes, and are thus clearly linked with the magnetic activity cycle (*cf*. Tuominen et al., 1983; Zhao & Kosovichev, 2004; Rempel, 2007; McIntosh et al., 2014).  A relationship between the zonal flows of the torsional oscillations and the solar meridional flows is also frequently seen (Tuominen et al., 1983; Bogart, 1987; Zhao & Kosovichev, 2004; Cameron and Schüssler, 2010; Švanda et al., 2010; Komm et al., 2011, 2015; Zhao et al., 2014).  The existence of a relationship linking the variability of the zonal and meridional flows suggests that both forms of variability may possibly arise due to a common mechanism.  However, the origins of the torsional oscillations, and their time variability, remain obscure (Rempel, 2007); this has been recognized as a key open question for future investigation (*cf*. Cameron et al., 2016).

While the torsional oscillations appear to be too weak to significantly influence the Sun's magnetic activity (Cameron et al., 2016), there is evidence to indicate that they (quite remarkably) extend as coherent flows throughout the depth of the convection zone (Howe et al., 2000, 2005; Antia & Basu, 2000; Verontsov et al., 2002; Howe, 2009).  Clearly the rotational variability of the Sun, as evidenced by the torsional oscillations, is intimately connected with the solar magnetic cycle and hence with the underlying mechanism(s) of the solar dynamo.

If orbit-spin coupling is operative in the case of the Sun, then we might reasonably expect to find some relationship linking the variability of the solar rotation with the identified forcing function for the coupling (*dL/dt*).   Such a relationship has already been reported (Juckett, 2003), for the torsional oscillations, in a study of sunspot group data extending from 1874 to 1999.  Additionally, in a separate study, Javaraiah (2005) found a significant positive correlation of the solar equatorial rotation rate with *dL/dt* for the interval from 1879 through 1945, which corresponds fairly closely to the time interval illustrated in the lower panel of Fig. 1.  These two studies provide *direct observational evidence of a coupling of the orbital and rotational motions* *of the Sun*.  Wilson et al. (2008) and Wilson (2013) provide additional evidence of a dynamical coupling of solar rotation speeds and orbital motions.

We now turn our attention to a consideration of the more complex phase relationships of the solar magnetic cycle and *dL/dt* for the years from 1950 to date.

### *4.2. Solar motion and solar activity, 1950-2016*

In order to understand the origins of the differences between this time period and those discussed earlier (i.e., with reference to Fig. 1), we must briefly step back and consider the nature of the barycentric orbital revolution of the Sun in slightly more detail.  Polar plots illustrating the orbital motion of the center of the Sun during two intervals of interest (i.e., from 1888-1948, as in Fig. 1, and from 1950-2030) are provided in Fig. 3.  For the years from 1888-1948, in panel a), smaller and larger orbital loops alternate to form a relatively regular and symmetric pattern, which has been labeled a 'trefoil' (Charvátová & Střeštík, 1991).  Panel b) of Fig. 3 illustrates the solar motion for the period from 1950-2030.  The azimuthal symmetry exhibited in panel a) is not in evidence here.  We note that in panel b) the path of the center of the Sun more closely approaches the solar system barycenter than is the case in panel a), while at other times the radial displacements in b) are seen to be larger than is the case at any time during the prior interval. These differences are important, as the barycentric angular momentum *L* of the solar motion depends on both the distance and the velocity.  Unlike the situation in Keplerian orbital motion, the velocity of the Sun is smallest when the Sun approaches the barycenter most closely, and is largest when farthest away.  Thus the range of variability of *L* is considerably larger in panel b) than in panel a).   To highlight the significant differences in the nature of the solar motion between these intervals, the regular and well-ordered conditions of panel a) have been characterized as the 'Sunday Driver' mode (Shirley & Duhau, 2010), while the less ordered and more variable conditions corresponding to panel b) have in contrast been referenced as the 'Teenage Driver' mode.

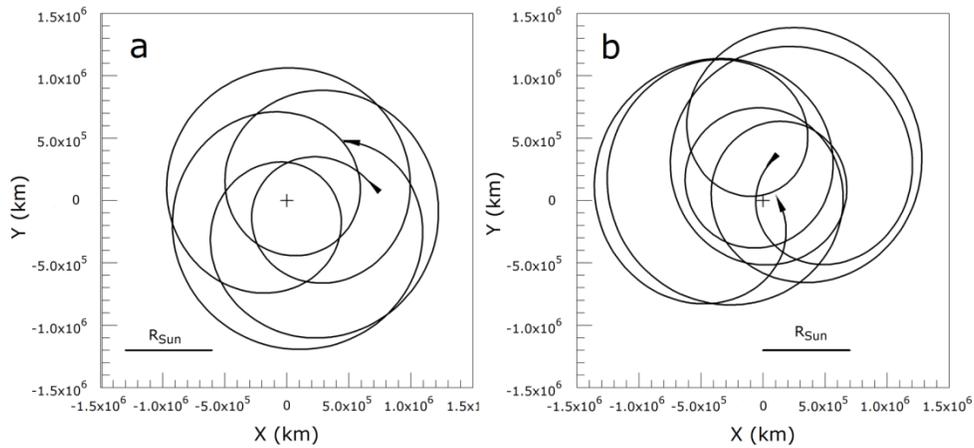

**Figure 3**. Polar plots illustrating the motion of the center of the Sun with respect to the solar system barycenter (+) for two intervals considered in this paper. The radius of the Sun is shown in each panel to provide scale. Motion is plotted relative to the J2000 ecliptic coordinate frame. a): 1888-1948 (corresponding to the 60-year time period of the lower panel of Fig. 1). Smaller and larger loops alternate to form a relatively regular and symmetric pattern, which has been labeled a 'trefoil' (Charvátová & Střeštík, 1991). b): 1950-2030 (corresponding to the time interval of Fig. 4 below). An azimuthally asymmetric pattern is formed, with the Sun traveling both much closer to the barycenter, and much farther from the barycenter, at different times, in comparison with a).

The differences between the Sunday Driver mode and the Teenage Driver mode may be directly attributed to differences in the azimuthal distribution (in celestial longitude) of the outermost giant planets Uranus and Neptune. For extended intervals, as in panel 3a, when these slow-moving bodies are near solar opposition with respect to one another, their combined influence on the displacement of the Sun from the barycenter is much reduced. At other times, as in panel 3b, the interference of their contributions typically leads to a larger range of variability of the Sun's radial distance ($R$) and orbital angular momentum $L$. The solar motion is described more completely in the studies by Jose (1965), by Fairbridge and Shirley (1987), by Charvátová & Střeštík (1991), and by Shirley (2015).

Figure 4 illustrates the waveform of the orbit-spin coupling forcing function $dL/dt$ for the years 1950-2030, in juxtaposition with the Hale cycle sunspot numbers, in the same format as in Fig. 1. Note however that the vertical scale has been expanded in Fig. 4 due to the greater range of $dL/dt$ during this period. We immediately recognize that the correspondence in phase of the forcing function and the 22-year magnetic activity cycle seen in Fig. 1 is largely absent for these years. The polarity (or sign) of the $dL/dt$ waveform switches rapidly in the low-amplitude cycles near 1970 and 2012. (Both of these intervals correspond to times of close approach between the Sun and the solar system barycenter, as illustrated in Fig. 3b). Near the center of the figure is a rapid oscillation of the $dL/dt$ waveform that represents a rare episode of retrograde solar motion and negative solar orbital angular momentum. Javaraiah (2005) has linked such intervals with past violations of the Gnevyshev-Ohl rule (the 'even-odd' cycle rule). This episode of retrograde solar motion may be identified in Fig. 3b as the loop located just above the arrowhead denoting the end of the trajectory. This loop falls short and fails to complete a prograde (counterclockwise) orbit around the barycenter.

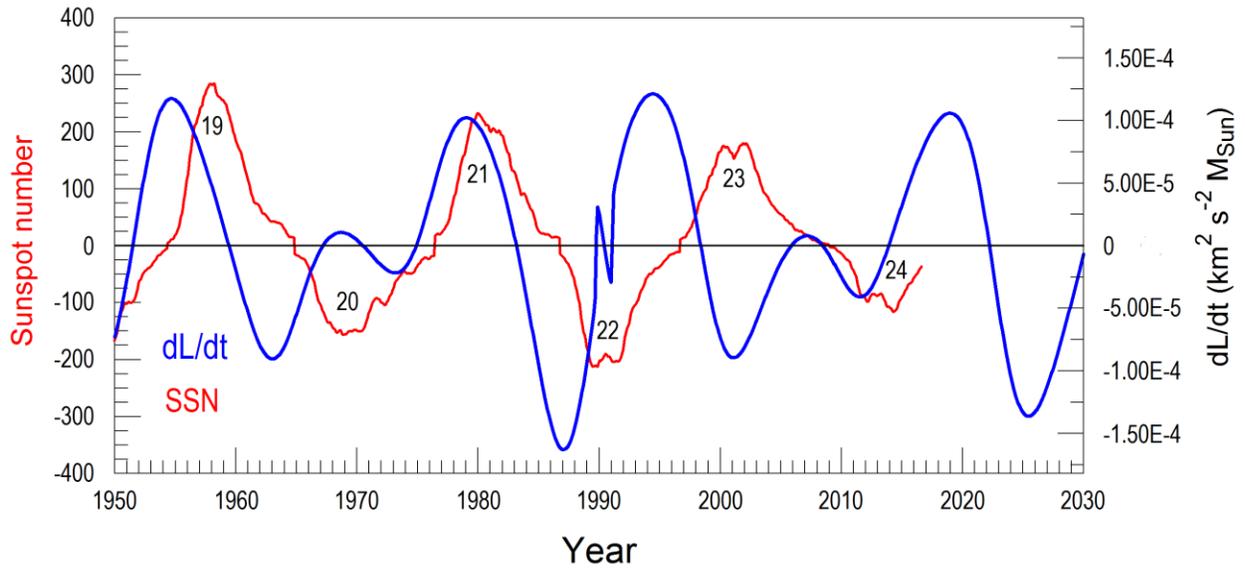

**Figure 4**. SILSO monthly mean sunspot numbers (red) (Clette et al., 2014) and *dL/dt* (blue), for the years 1950-2030. Sunspot Cycle numbers (19-24) are also indicated. As in Fig. 1, only giant planet contributions are employed in the calculation of the *dL/dt*.

We begin by considering the two time periods of Fig. 4 in which the two waveforms are most nearly out of phase. This is seen to be the case during Sunspot Cycles 20 and 23. With reference to the discussion accompanying Fig. 1, and the predictions of the orbit-spin coupling hypothesis (Section 3), it is tempting to suppose that the out-of-phase relationship of the waveforms displayed may correspond to a condition of *destructive interference* between the putative forcing function *dL/dt* and the dynamo mechanism for these cycles. In the case of Cycle 20, relatively weak out-of-phase forcing is implied, as indicated by the small amplitude of the positive polarity interval of *dL/dt* during that cycle. We see that the peak amplitude of the sunspot numbers in Cycle 20 was significantly lower than that for the preceding or following cycles. While the diminished amplitude of the sunspot maximum of Cycle 20 has occasionally been noted in the literature, the more recent events and phenomena of Cycle 23, considered next, have been investigated in much greater detail.

### *4.3. Events and phenomena of Sunspot Cycle 23*

Sunspot Cycle 23 was noteworthy for its length and its long, drawn-out minimum, with very few sunspots observed (Russell et al., 2010). In addition, dramatic reductions (approaching 50%) in the strength of the solar magnetic field and particularly the Sun's polar fields were observed (Dikpati et al, 2004; Schatten, 2005; Wang et al., 2009; Dikpati et al., 2009; Janhardhan et al., 2010). The associated changes in the heliospheric magnetic field (Smith & Balough, 2008; Janardhan et al., 2011) and the solar wind (Russell et al., 2010; Jian et al., 2011; Janardhan et al., 2011) were unprecedented in the space age. Complex variations of meridional flows were observed (Basu & Antia, 2010; Jiang t al., 2010; Komm et al., 2011, 2015a). A number of attempts to account for these observations through modeling with variable meridional flow speeds were made (Dikpati et al., 2004, 2010; Wang et al., 2009; Janardhan et al., 2010; Basu & Antia, 2010; Nandy et al., 2011). However, some uncertainty still attends this topic (*cf.* Cameron & Schüssler, 2010), and it is fair to say that the question of how differences in meridional flow speeds may have led to the weakening of polar fields in Cycle 23 remains open (Dikpati, 2011; Wang, 2016).

Cycle 23 was anomalous in other ways. The Sun's equatorial rotation rate was nearly constant throughout the cycle, in a marked departure from prior cycles (Javaraiah et al., 2009); at the same time, unusually fast meridional motions of sunspots were also seen (Javaraiah, 2010). Marked hemispheric asymmetries were also noted (*cf.* Zaastri et al., 2006; Sindhuja et al., 2014; Ravinda & Javaraiah, 2015; Belucz et al., 2015; Wang, 2016; Gopalswamy et al., 2016). Finally,

Haber et al. (2002) describe the emergence and development of a large submerged meridional flow cell with reversed circulation in the Sun's northern hemisphere in the years 1998-2001. This appears to have played an important role in the subsequent evolution of polar fields in the later years of the cycle (Dikpati et al., 2004; Wang, 2016). A number of other observations of "countercells" have been reported (González Hernandez et al., 2006; Ulrich, 2010), but some other studies find no evidence of these (Hathaway & Rightmire, 2010; Rightmire-Upton et al., 2012; Komm, et al., 2015b). In some cases, the reported counterflows may have been artifacts due to small errors in the astronomical elements of the solar rotation axis that were employed in observation processing (Komm et al., 2015b).

In advance of numerical modeling, we cannot with confidence associate any of the above phenomena with the out-of-phase condition of the waveforms of Fig. 4 during Cycle 23. However, an observational record of meridional flow speeds of small magnetic features spanning the duration of Cycle 23 is available (Hathaway & Rightmire, 2010). That sample has been employed for hypothesis testing.

## 5. SOLAR MERIDIONAL FLOWS AND $dL/dt$, 1996-2009

Meridional accelerations largely dominate in equatorial and mid-latitudes in the *CTA* acceleration field of Fig. 2. Further, the time-variability of the *CTA* field is driven by the variability of $dL/dt$, as displayed in Figs. 1 and 4. Following the discussion of Paper 1 and Section 3 above, we are able to formulate an initial working hypothesis, as follows: Orbit-spin coupling accelerations may effect a modulation of meridional flow speeds on and within the Sun.

*5.1. Time series employed*

To test this hypothesis, we employ the record of meridional flow speeds of small magnetic features on the Sun of Hathaway & Rightmire (2010). These authors analyzed observations from the Michelson Doppler Imager (MDI) on board the Solar and Heliospheric Observatory (SOHO) spacecraft spanning a period from May 1996 to June 2009. Data from active regions was excluded from the analysis. Details of the observations and data and error analyses are provided in Hathaway & Rightmire (2010, 2011). Meridional flow speeds were averaged over time periods corresponding to one Carrington rotation, leading to a data set consisting of 167 points (only 7 Carrington rotations from the period studied are not represented in the dataset). The SOHO/MDI data for CR 1909 through CR 2085 are displayed in the middle panel (b) of Fig. 5.

For purposes of comparison we employ high-resolution JPL Horizons ephemeris data (Giorgini et al., 1996; Georgini, 2015) for the position and velocity of the Sun, obtained using a time step of one day, for the same years. We subsequently process this data to obtain the $d\boldsymbol{L}/dt$ waveform as described in the Appendix. As with the SOHO-MDI meridional flow speeds, we obtain time averages of this parameter corresponding to the interval of a single Carrington rotation. The resulting $d\boldsymbol{L}/dt$ waveform is displayed in panel (a) of Fig. 5.

$d\boldsymbol{L}/dt$ is represented in Fig. 5a in three different ways. The underlying smooth curve (in black) represents giant planet contributions only, as in Figs. 1 and 4. The blue curve represents the waveform with contributions from all the planets, at full temporal (1-day) resolution. The superimposed orange symbols represent our averaged data, with each point corresponding to one Carrington rotation (CR).

The higher-frequency modulation of the $dL/dt$ waveform evident in Fig. 5a is due to the influence of the inner planets. As first noted by Wood & Wood (1965), and as later shown in Shirley et al. (1990), the inner planet contributions become more important for the higher time derivatives of dynamical quantities. Synodic periods of Mercury and the giant planets, and also Venus and the Earth and the giant planets, may be extracted by spectral analysis from series such as that of Fig. 5b. The displacements of the Sun's center due to the inner planets are too small to be easily seen on the scale of Fig. 3, but this does not mean they are non-existent. While we have thus far neglected the inner planet contributions, for completeness in the following comparison it is of interest to include these.

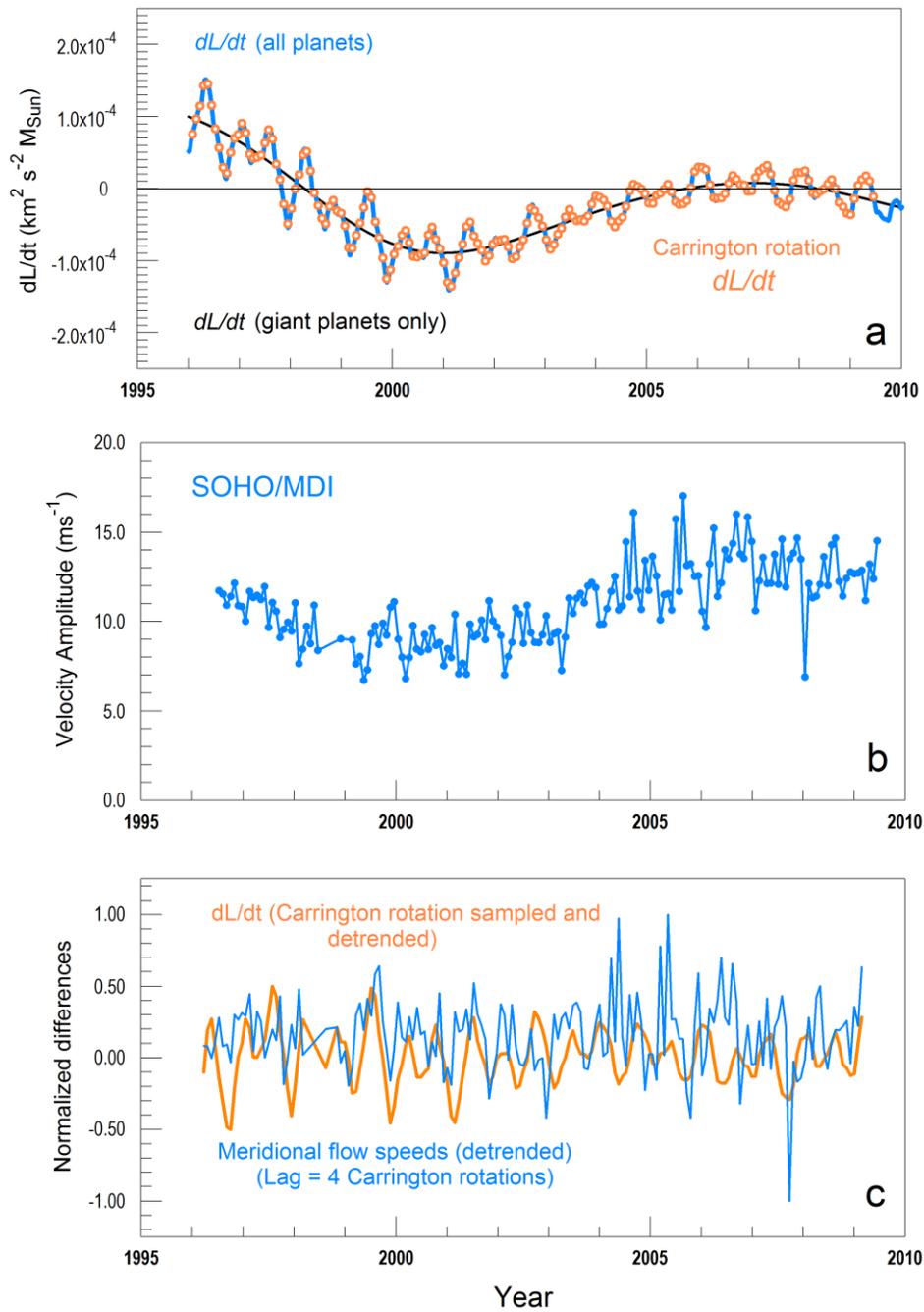

**Figure 5**.  (a):  The putative dynamical forcing function *dL/dt*. (b):  SOHO/MDI time series of solar meridional flow speeds of small magnetic features (Hathaway & Rightmire, 2010).  (c): Comparison of normalized, detrended time series of meridional flows and *dL/dt* (with meridional flows lagged by 4 CR).  The probability for the lag-4 correlation (with 165 d.f.) is *p*=0.011.

*5.2. Statistical comparisons*

We first search for a correlation between the time series of Fig. 5a ($dL/dt$) and the series for the measured meridional flow speeds of Fig. 5b. We compare the Hathaway & Rightmire (2010) series (HR2010) both with the giant-planets-only $dL/dt$ waveform and with the waveform representing $dL/dt$ with all planets considered, in order to assess the possible role and effects of the higher-frequency modulation due to the inner planets. We also address this latter question by means of a second comparison, this time comparing detrended versions of the HR 2010 and all-planets $dL/dt$ series. For this comparison the series were each detrended by the subtraction of a $4^{th}$ order polynomial fit to the data. The resulting detrended series are shown in the bottom panel of Fig. 5.

Dikpati and Anderson (2012) assessed the response time of a flux-transport dynamo to changes in solar meridional flow speeds. Their simulations indicated that the response time to changes in flow speeds was 4-6 months, over a range of flow perturbation conditions. The following comparison was added to the present experiment in order to explore the possible implications of the results of Dikpati & Anderson (2012). For each of the 3 comparisons specified above, we also calculate lag-correlation coefficients, employing lags (of the meridional flows series relative to $dL/dt$) of 1 to 6 Carrington rotations. The results are provided in Table 1.

Considering first the comparison between the giant-planets-only dynamical waveform and the HR2010 series, we find that the highest linear correlation ($r$=0.52) is obtained with a lag of 0 months. While the statistical significance of all the comparisons in the third column is high ($p < .001$, significance levels $> 99.9\%$), we see that the value of $r$ drops with each increment in the lag time. The significance levels remain high because of the relatively small shift in time introduced (only 6 Carrington rotations out of 174 for the full period).

We observe a different behavior in the comparison of the all-planets *dL/dt* series with the HR2010 meridional flow speeds. Here the highest value of the correlation coefficient is obtained at a lag of 3 months, with lesser correlations obtained for both lower and higher lag times. The uniformly high significance levels obtained for all comparisons in the fifth column of the Table is most likely due to the strong correlation of the two series at low frequencies (as found in the giant-planets only test). A comparison of detrended time series is thus called for.

| Lag | HR 2010 : Giant Planets | | HR 2010 : JPL Horizons | | HR 2010 : JPL Horizons (Detrended) | |
|---|---|---|---|---|---|---|
| (Carrington rotations) | *r* | *p* | *r* | *p* | *r* | *p* |
| | | | | | | |
| 0 | **0.520** | < 0.001 | 0.438 | < 0.001 | 0.060 | 0.221 |
| | | | | | | |
| 1 | 0.503 | < 0.001 | 0.436 | < 0.001 | 0.015 | 0.432 |
| 2 | 0.485 | < 0.001 | 0.443 | < 0.001 | 0.056 | 0.236 |
| 3 | 0.468 | < 0.001 | **0.458** | < 0.001 | 0.159 | 0.020 |
| 4 | 0.449 | < 0.001 | 0.451 | < 0.001 | **0.177** | **0.011** |
| 5 | 0.430 | < 0.001 | 0.415 | < 0.001 | 0.105 | 0.088 |
| 6 | 0.415 | < 0.001 | 0.376 | < 0.001 | 0.036 | 0.322 |

**Table 1**. Linear correlation coefficients and corresponding probabilities for comparisons of *dL/dt* with meridional flow speeds. The first column lists applied time lags (ranging from 0 to 6 CR) of the HR2010 series (Hathaway and Rightmire, 2010) with respect to *dL/dt*. For all comparisons, the number of data points is 167 and the degrees of freedom is 165. For all comparisons, data points represent averages over a single Carrington rotation. The largest value of the correlation coefficient *r* for each set of comparisons is highlighted. The probability *p* for obtaining the observed correlation coefficient is also provided.

Linear correlation coefficients for comparisons of the detrended series are likewise provided in Table 1. Substantial levels of random motions are present in the HR 2010 time series (*cf*. Section 7 of Hathaway & Rightmire, 2011). This noisy component may be expected to negatively impact the outcomes of our statistical comparisons. Nonetheless we detect a systematic pattern (first increasing, then decreasing) in the correlation coefficients for the lagged comparisons. The highest correlation ($r$=0.177) is found with a lag of 4 CR. The comparisons for lags of 3 CR and 4 CR achieve statistical significance at the 98% level or better. Figure 5c illustrates both series, with lag of 4 CR imposed.

The results compiled in Table 1 document a strong correlation linking observed meridional flow speeds of small magnetic features on the Sun with the variability of the dynamical forcing function *dL/dt* during Sunspot Cycle 23. In our comparisons of the full resolution (all-planet contributions) series, the best correlation is found for a lag of 4 Carrington rotations, which is, perhaps coincidentally, in substantial agreement with the results of the simulations by Dikpati & Anderson (2012), who found dynamo response times of 4-6 months to changes in imposed meridional flow speeds.

## 6. RECENT SOLAR VARIABILITY AND ORBIT-SPIN COUPLING (2):

### *6.1. Observations during Sunspot Cycle 24*

Cycle 24 is presently in its decaying phase, having attained a maximum smoothed sunspot number of ~80 in April, 2014 (Hathaway, 2015). This represents the lowest sunspot cycle maximum recorded in the past century. Altrock (2011) has described a "historically long and weak start" to Cycle 24, using coronal Fe XIV emission data. Howe et al. (2013) detailed a number of anomalous observations of the torsional oscillation pattern during the rise of Cycle 24, and inferred that a slowing of the solar rotation rate at mid- to high latitudes may have occurred. Zhao et al. (2014) described strong hemispheric asymmetries and temporal variations in the torsional oscillations.   Unusual polar conditions, including pronounced north-south asymmetries in reversal processes, are described by Sun et al. (2015) and Gopalswamy et al. (2016). Observations made during the rising phase of Cycle 24 thus appear to be consistent with a continuation of the anomalous behaviors and weakening trend of the solar dynamo and magnetic field that were prominent aspects of the preceding cycle.  However, Sheeley & Wang (2015) have recently noted a sudden strengthening of the Sun's large-scale magnetic field, occurring in the second half of 2014.  This was attributed not to an increase in sunspot activity, but to an increase in the emergence of flux in active regions.  With reference to the plotted $dL/dt$ and sunspot cycle waveforms of Fig. 4, we are able to offer the following speculative interpretation.

A fundamental difference between Cycles 23 and 24, as illustrated in Fig. 4, lies in the relative phasing of the waveforms representing $dL/dt$ and the solar (Hale) sunspot cycle. Following the 2009 transitional interval of the $dL/dt$ waveform, at the end of Cycle 23, we note that both waveforms have returned to an in-phase relationship, similar to the relationships illustrated in Fig. 1, but lasting only for about 3 years.  Termination of the in-phase interval occurs in 2012, shortly before the first peak of the Cycle 24 sunspot numbers was attained. Following a short dip, the sunspot numbers continued to increase, attaining their peak values in 2014.  By this time, the $dL/dt$ waveform had returned to transitional (near zero) values.

Under the present hypothesis, the rejuvenation of the solar magnetic field in 2014 may be interpreted as a consequence of the constructive, in-phase forcing that occurred during the first 3 or so years of Cycle 24.  The continued rise in sunspot activity for a small number of years following the attainment of peak *dL/dt* forcing might either be attributed to an "overshoot" phenomenon, or to the influence of some delay introduced by an unknown system memory effect.  In either case, this feature of Cycle 24 is in some ways similar to features seen in Fig. 4 in association with Sunspot Cycles 19 and 21, where peaks of the sunspot numbers curve likewise follow the peaks of the *dL/dt* waveform by a small number of years.

### 6.2. Anti-correlation of meridional flow speeds and sunspot cycle activity levels

There may be more than one explanation, under the orbit-spin coupling hypothesis, to account for cases of anti-correlation of meridional flow speeds and sunspot activity levels during recent cycles.  In the foregoing, we have interpreted the observed slower flow speeds near the sunspot cycle maximum of Cycle 23 as possibly being due to some thus-far unspecified destructive interference effect.  (We will explore one possibility for an underlying mechanism below in Section 7).   We immediately recognize, however, that other similar episodes of anti-correlation have occurred, and that not all of these have exhibited the same phase relationship as that of Cycle 23.

Work by Komm et al. (1993) revealed a similar anti-correlation, of flow speeds and sunspot activity levels, for the years 1970-1990.  Consider the waveforms of Fig. 4 for the time of the sunspot minimum lying between Cycles 21 and 22, in late 1986.  The sunspot minimum coincides with an extreme (negative) value of the *dL/dt* waveform.  The following sunspot maximum, of Cycle 22, in late 1989, on the other hand, was coincident in time with a zero crossing of the *dL/dt* waveform (albeit an unusual one, associated with an episode of retrograde motion).

If, as stated in our working hypothesis of Section 5, meridional flow speeds are to some extent modulated directly by orbit-spin coupling accelerations, then it may be possible to interpret the fast meridional flow speeds of 1986 as likely resulting from an intensified circulation, forced by a high level of *CTA* acceleration, accompanying the extreme values of $dL/dt$. Conversely, the low flow speeds in late 1989 might be reasonably be attributed to circulatory relaxation, reflecting the minimal levels of forcing provided by the *CTA*, during this transitional interval.   In this scenario, the waveforms of Fig. 4 are not found to be in opposition (as in Cycle 23), but are approximately in quadrature.  Thus here a completely different interpretation of the negative correlation of flow speeds and sunspot activity can be offered.

Let us once more consider the implications for meridional flow speeds of the in-phase 'Sunday Driver' episodes of Fig. 1 and the out-of-phase 'Teenage Driver' cases of Fig. 4.  In the Sunday Driver case, with cycles of circulatory intensification coinciding with the peaks of the $dL/dt$ waveform, we might expect to find a *positive* correlation of flow speeds and sunspot cycle activity levels, rather than the anti-correlation that has been found in recent decades.  The negative correlations observed in recent cycles thus may not necessarily represent an immutable rule, over much longer time intervals.  A direct relationship of flow speeds and sunspot activity levels should not be summarily excluded, discounted or otherwise dismissed from consideration in future modeling activities, due solely to the negative correlations found for recent cycles.

In connection with the strong phase and amplitude modulation of the $dL/dt$ waveform in the Teenage Driver mode, as displayed in Fig. 4, and its possible relationship to solar dynamo excitation:  A cogent interpretation for the observations can be well summed up by the following statement, which we have borrowed from Passos & Lopes (2008):  "It can be interpreted as if the system is constantly re-adjusting to an equilibrium solution that stubbornly refuses to set still."

## 7. DISCUSSION

Our investigation has uncovered a direct relationship linking the variability of observed solar meridional flow speeds with the time rate of change of the solar orbital angular momentum $dL/dt$ during Sunspot Cycle 23. The relationship observed is consistent with expectations based on the physical hypothesis of orbit-spin coupling described in Paper 1. The production of constructive and destructive interference effects within solar and planetary atmospheric circulations is an expected consequence of the specific form of orbit-spin coupling formulated in Paper 1.

The physical hypothesis of P1 predicts cycles of intensification and relaxation of circulatory flows, whose timing is linked with the phase of the $dL/dt$ waveform. An important feature of this hypothesis is that the timing of the predicted pulsations of large-scale flow speeds need not maintain any consistent relationship with the phase of the magnetic activity cycle (as in Fig. 4). We have cited this aspect above in our interpretation of the results of Komm et al. (1993). We suspect that this characteristic may also be relevant to questions of the evolution and development of polar fields in different cycles. The time intervals following the polarity reversal of the Sun, shortly after sunspot cycle maxima, might either be characterized by conditions of flow speed intensification, of two opposing signs, or by flows experiencing much reduced acceleration, during transitional intervals. We have not yet attempted to resolve possible real-time relationships between polar fields observations and the forcing function. Numerical modeling may better address this problem.

While the orbit-spin coupling hypothesis of Paper 1 generally appears to provide a useful conceptual framework for understanding some features of the solar magnetic cycle, a number of key open questions from Section 1 have not yet been considered. Among these are the issues of the source(s) of system memory for the dynamo, and the recent observations of multiple solar meridional flow cells, some with reversed circulations, and with lifetimes of some years. Our interpretation of the events and phenomena of Sunspot Cycle 23 posits some as-yet unspecified form of destructive interference between the (putative) driven flows and the underlying magnetic dynamo process. Can we conceive of a specific mechanism that may address this suite of questions, and which may be at the same time susceptible to evaluation and testing through numerical modeling?

### 7.1. A possible mechanism for angular momentum sequestration and redistribution

The dynamical waveforms of Figs. 1 and 4 represent and describe an ongoing transfer and exchange of orbital angular momentum between the Sun and planets. The Sun gains orbital angular momentum during intervals when the sign of the *dL/dt* waveform is positive, and yields up orbital angular momentum when it is negative. The fundamental premise of the orbit-spin coupling hypothesis is that some portion of the orbital angular momentum thus exchanged may be temporarily deposited within, or withdrawn from, the rotating subject body involved (in this case, the Sun). The acceleration field defined by Equation (1) and illustrated in Fig. 2 exhibits dominantly meridional accelerations in equatorial and middle latitudes. Taken together, these considerations lead us to suspect that meridional flow cells within the Sun may either be formed, or strengthened, in response to the persistent presence of the predicted meridional accelerations, during intervals with favorable phasing, such as those of Fig. 1. If so, then these cells are likely to represent repositories for the temporary storage of deposited angular momentum. Additionally, when the putative driving accelerations reverse direction, or otherwise modify the large-scale flow, we may suppose that a form of destructive interference with the previously established meridional cellular circulation may be engendered. That is, an altered large-scale circulation may possibly act to retard and actively spin-down previously loaded meridional flow cells, possibly through processes similar to those proposed to operate near and within the solar tachocline.

Some additional insight may be gained by a further consideration of the *CTA* acceleration field illustrated in Fig. 2. In the northern hemisphere, at the left side of the figure, the northward-directed accelerations will presumably act to enhance poleward flows of solar materials at the surface. However, at depth, the velocities of the corresponding return flows would instead be retarded. The magnitude of the acceleration retarding the return flow would, however, be somewhat smaller (due to the dependence of the acceleration magnitude on the radius vector $r$ in Equation (1)). The opposite situation would prevail in the southern hemisphere, at the left side of Fig. 1. As previously noted in connection with Fig. 2, hemispheric asymmetries of meridional flows might plausibly arise in this manner.

The processes envisioned above are likely to operate (and produce cascading effects) on multiple timescales, including scales shorter than or comparable to the period of one solar sidereal rotation. We suspect that their cumulative effects may reflect the longer-term evolution of the putative forcing function $dL/dt$. Once again, numerical modeling is likely to be required to gain traction on this question.

The above-described angular momentum storage and conversion process could potentially provide a significant component of system memory for the solar dynamo; and it has the virtue of being, at least in principle, verifiable through numerical modeling.

## 7.2. Recommendations for future work

The algorithms provided in the Appendix allow the calculation of orbit-spin coupling accelerations for the Sun as a function of both space and time. Incorporation of the *CTA* within the dynamical core of the MarsWRF GCM was accomplished without great difficulty (Mischna & Shirley, 2017). We anticipate that the inclusion of the *CTA* within existing dynamo models may present a somewhat different set of problems, for instance, in connection with specifying the angular velocity of axial rotation $\omega_a$ as a function of depth and latitude within the Sun, and in accounting for backreaction on the driven flows due to magnetic forcing. Clearly 3-dimensional models (*cf*. Miesch & Dikpati, 2014; Passos et al., 2015, 2016; Beaudoin et al., 2016; Guerrero et al., 2016) may have some advantages, in light of the 3D structure of the acceleration field, as depicted in Fig. 2. Consideration should not be limited to flux transport dynamo models, as other approaches may offer advantages for certain problems.

A wide range of experiment designs are possible. Observations of a wide variety of solar phenomena may be employed for sanity checking of model outcomes. Earlier in this paper, an initial approach to modeling was described, in which the coupling efficiency coefficient $c$ was to be iteratively modified, beginning with small values, and continuing until an acceptable value might be found (as determined by a criterion of obtaining improved agreement of model outcomes with observations).

Investigations that constrain the permissible values of *c* for the solar case may have implications that extend far beyond the solar realm.  As discussed more fully in Paper 1, it is of considerable interest to learn whether nature prefers, and widely employs, a non-zero value of *c*.  There are many cases in the universe where orbital interactions are more extreme than those of our solar system.  In our standard models, in both Newtonian and relativistic contexts, orbital and rotational motions are considered to be independent and uncoupled, aside from certain small effects (including precession-nutation, and tidal friction) that arise due to tidal gradients of gravitational fields (see the Appendices of P1).  Paper 1 introduces a new and potentially far more effective type of coupling, which may be operative at much greater distances than is the case with tidal processes.

While angular momentum conservation is assumed throughout our presentation, dissipative processes involving losses of orbital and rotational energies, similar to those predicted in studies of tidal friction, are nonetheless envisioned (P1).  The dynamo mechanism fundamentally involves a dissipative conversion of the kinetic energy of plasma motions to electromagnetic energy.  The Sun's present rotation rate, perhaps coincidentally, is in the lowest 10% of values determined for similar G-2 stars (Blizard, 1981).  If orbit-spin coupling leads to a secular spin-down of the solar rotation rate, then momentum conservation would presumably dictate a corresponding secular transfer of momentum to the linked 'reservoir' of the orbital motions of the planets of the solar system.

### 7.3. The 'chronometer' within the Sun

R. H. Dicke (1978), writing nearly 40 years ago, determined that "the sunspot cycle shows no statistical indication of a random walk in phase," and concluded that the sunspot cycle

"seems to be paced by an accurate clock inside the Sun." The question of what sets the dynamo period is still recognized as an outstanding open question of solar physics (Charbonneau, 2014; Cameron et al., 2016). Cyclic behavior has been obtained from disparate dynamo models, with a variety of adjustable parameters. Nonetheless, as recently noted in Cameron et al. (2016), "we have no clear basis, in either observations or theory, for understanding why… the dynamo period is 11 years."

A periodicity of ~22.4 years, virtually identical to that of the Hale cycle, arises naturally from combinations of the orbital periods characterizing the system of the four giant planets (Bureau & Craine, 1970; Mörth & Schlamminger, 1978). On the basis of the correlations displayed in Figs. 1 and 5, and the many other relationships and correspondences found linking the dynamical forcing function $dL/dt$ with solar phenomena, we conclude that the 22-year Hale cycle of the solar dynamo is most likely driven by the same solar system dynamical processes that determine the trajectory of the Sun's orbital motion about the barycenter of the solar system. We are unaware of any other physical process that may similarly act as a pacemaker for the Hale cycle. The orbit-spin coupling hypothesis described here and in Paper 1 supplies a testable and fully deterministic physical mechanism to account for the system behaviors observed.

## 8. CONCLUSIONS

This paper has outlined the implications for solar physics of a newly-formulated physical hypothesis describing a weak coupling of the orbital and rotational motions of extended bodies. The hypothesis is susceptible to direct evaluation and testing by numerical modeling.

Algorithms for calculating the predicted orbit-spin coupling accelerations are provided in the Appendix to this paper.

A statistical test of the orbit-spin coupling hypothesis was performed in this investigation. The dynamical time series of the rate of change of the Sun's orbital angular momentum $dL/dt$ was compared with a time series representing measured solar surface meridional flow speeds of small magnetic elements during Sunspot Cycle 23 (Hathaway & Rightmire, 2010). We obtained correlation coefficients of ~0.4 to 0.5 which are statistically significant at the 99.9% level. Our investigation has thus uncovered a direct relationship linking the variability of observed solar meridional flow speeds with the time rate of change of the solar orbital angular momentum $dL/dt$ during Sunspot Cycle 23. The relationship observed is consistent with expectations based on the physical hypothesis of orbit-spin coupling described in Shirley (2017).

A new conceptual model for the role of meridional flow cells as a component of the solar dynamo has been introduced. Meridional flow cells are here proposed to constitute repositories for the temporary storage of angular momentum deposited within the rotating Sun as a consequence of orbit-spin coupling. This role is not inconsistent with our present understanding of the solar dynamo; we have long recognized the presence of an intimate relationship linking the operation of the solar dynamo with variations of the solar rotation, and with variations of solar meridional flows. The contribution of the present study is found in the proposed weak coupling between the motions of the solar rotation and the solar barycentric revolution.

A strong motivation for follow-on numerical modeling investigations is found in the potential benefits to society of an improved capability to forecast the future course of solar variability. The dynamical forcing function $dL/dt$ may be calculated with great accuracy for period of many hundreds of years in the past and in the future. An improved understanding of the solar dynamo mechanism could lead to an improved understanding of the nature and frequency of extreme space weather conditions in the near-Earth environment of the solar system.

APPENDIX: ALGORITHMS FOR CALCULATING THE ORBIT-SPIN COUPLING

ACCELERATIONS

We first obtain the instantaneous orbital angular momentum of the Sun with respect to the solar system barycenter, using the following equation, from Jose [1965];

$$\boldsymbol{L} = [(y\dot{z} - z\dot{y})^2 + (z\dot{x} - x\dot{z})^2 + (x\dot{y} - y\dot{x})^2]^{1/2} \tag{A1}$$

Here the required quantities are the positional coordinates ($x$, $y$, $z$) and velocities ($\dot{x}$, $\dot{y}$, $\dot{z}$) of the subject body with respect to the solar system barycenter. The mass is not explicitly included (but must be supplied later as a multiplicative factor for quantitative comparisons). The positions and velocities required may be obtained from JPL's online Horizons ephemeris system [Giorgini et al., 1996; Giorgini, 2015]. To obtain the time derivative $d\boldsymbol{L}/dt$ it is simplest to merely difference the values of each of the vector components of $\boldsymbol{L}$ for two adjacent times, and divide each resulting component by the time difference. We then assign an intermediate time value to the 3-component Cartesian rate of change vector obtained.

The time derivative of the solar orbital angular momentum $d\boldsymbol{L}/dt$ obtained in this way will be referenced to the coordinate system specified when accessing the JPL Horizons system (the J2000 ecliptic system was employed for the calculations of Section 3 of this paper). As a practical matter, for computational efficiency, it may be desirable to next perform a coordinate transformation or set of transformations to resolve the vector components of $d\boldsymbol{L}/dt$ in the "native" frame of the dynamo model employed. (In the calculation of Section 3, we did not perform this step, as the components of the rotational angular velocity were there obtained in the ecliptic frame).

The coordinate frame employed for the dynamo calculations will presumably most often be a Cartesian system with the $z$ axis aligned with the rotation axis of the Sun. In this frame, the (vector) angular velocity of rotation $\boldsymbol{\omega}_\alpha$ will have zero magnitude in the $x$ and $y$ directions. Numerical values for the astronomical elements of the rotation axis of the Sun (*cf.* Beck & Giles, 2005) will be required for this transformation of coordinates.

The next step is to form the cross product of $d\boldsymbol{L}/dt$ with the local (vector) angular velocity of rotation, $\boldsymbol{\omega}_\alpha$, which we emphasize must be represented *in the same coordinate frame*. Because the angular velocity of rotation, $\boldsymbol{\omega}_\alpha$, varies as a function of latitude and depth within the Sun, this calculation may be somewhat more involved than is the case for the atmospheres of the terrestrial planets.

Having thus formed the cross product $\dot{\boldsymbol{L}} \times \boldsymbol{\omega}_\alpha$ in the native 'dynamo frame,' as a function of radius and latitude variables, a likely next step would be to repetitively obtain the cross product of this with the set of position vectors identifying the dynamo model's grid point locations within the body of the Sun.

As indicated within the accompanying paper, the accelerations thus calculated are quite large. The coupling efficiency coefficient $c$ should be coded as an adjustable scaling parameter, for initial investigations, allowing the user to determine a value for the accelerations that may suitably modify model outcomes, such that physically and observationally absurd or otherwise catastrophic cases may be avoided.

Because meridional flows in most kinematic flux transport dynamos may be specified externally (*cf.* Dikpati et al., 2014), we anticipate that existing models (i.e., models not incorporating the coupling term accelerations) may be suitably updated and modified to initiate with a specified flows profile, and thereafter allowed to evolve, with flow speeds dictated by the accelerations added at each time step.

# ACKNOWLEDGEMENTS

This paper is dedicated to the memory of Rhodes W. Fairbridge.  We thank D. H. Hathaway for supplying the SOHO/MDI dataset.  Critical comments and suggestions from M. A. Mischna and M. I. Richardson materially improved this presentation.  The author has in addition benefited greatly from earlier discussions with many individuals, including H. T. Mörth, I. Charvátová, D. A. Juckett, J. Javaraiah, S. Duhau, D. J. McCleese, and M. A. Mischna.  A preliminary version of the statistical investigation of this paper was previously presented at the 2010 American Geophysical Union "Meeting of the Americas" (Shirley & Duhau, 2010).  This work was done as a private venture and not in the author's capacity as an employee of the Jet Propulsion Laboratory, California Institute of Technology.